\documentstyle[twocolumn,prl,aps,epsf,epsfig,amsmath]{revtex} 

\newcommand{\bea}{\begin{eqnarray}} 
\newcommand{\eea}{\end{eqnarray}}

\font\cmss=cmss12  
\def\1{\hbox{{1}\kern-.25em\hbox{l}}} 
\def\bfZ{\relax{\hbox{\cmss Z\kern-.4em Z}}} 
 
\begin{document} 
 
\title{A next-to-leading order analysis of deeply virtual Compton scattering} 
\author{A.~Freund$^{1}$\thanks{Email address: andreas.freund@physik.uni-regensburg.de} and
M.~McDermott$^{2}$\thanks{Email address: martinmc@amtp.liv.ac.uk}} 
\address{$^1$Institut f{\"u}r Theoretische Physik, Universit{\"a}t Regensburg,  
D-93040 Regensburg, Germany} 
\address{$^2$Division of Theoretical Physics, Dept. Math. Sciences, University of Liverpool, Liverpool, L69 3BX, UK} 
\maketitle 

\begin{abstract} 
We present a complete, next-to-leading-order (NLO), leading-twist QCD analysis 
of deeply virtual Compton scattering (DVCS) observables, in the ${\overline {MS}}$ scheme, 
and in the kinematic ranges of the H1, ZEUS and HERMES  experiments.  We use a 
modified form of Radyushkin's ansatz for the input model for the generalized parton 
distributions. We present results for leading order (LO) and NLO for representative 
observables and find that they compare favourably to the available data. 
\end{abstract} 

\medskip 
 
Deeply virtual Compton scattering (DVCS) \cite{rad,ji,diehl,mue,van,cf,ffs,exp},
$\gamma^* (q) + p (P) \to \gamma (q') + p (P') $, is the
most promising\cite{foot1} process for accessing generalized parton distributions (GPDs) \cite{rad,ji,mue,bfm,pet,ffgs}  
which carry new information about the  
dynamical degrees of freedom inside a nucleon.  
GPDs are an extension of the well-known parton distribution functions (PDFs) appearing 
in inclusive processes and are defined as the Fourier transform of  
{\it nonlocal} light-cone operators sandwiched between nucleon states of  
{\it different} momenta, commensurate with a finite momentum transfer in the t-channel. 
These distributions are true two-particle correlation functions and contain,  
in addition to the usual PDF-type information residing in the
so-called ``Dokshitzer-Gribov-Lipatov-Altarelli-Parisi (DGLAP) region'' \cite{dglap},  
supplementary information about the distribution amplitudes of virtual
``meson-like'' states in the nucleon in the so-called ``Efremov-Radyushkin-Brodsky-Lepage (ERBL) region'' \cite{erbl}.  
We have recently presented a full numerical solution of the associated renormalization  
group equations at next-to-leading order (NLO) accuracy, for both the polarized and unpolarized distributions 
\cite{frmc1}. 
 
In this paper, we report on a full NLO, leading-twist QCD analysis carried out for 
unpolarized and singly polarized (with respect to the lepton) DVCS and its physical 
observables based on the DVCS factorization theorem\cite{rad,cf}. 
The precise technical details of this analysis are presented in considerably more 
detail elsewhere \cite{frmc1,frmc2,frmcbig}. DVCS and its observables are accessed experimentally  
in deep inelastic scattering experiments \cite{exp} using the leptonic process   
$e^{\pm} p \to e^{\pm} p  \gamma$. Hence, on the lepton level, DVCS interferes with the 
Bethe-Heitler (BH) process which has the same final state 
(for BH, the final state photon is radiated off either the initial or
final state lepton). The exact DVCS kinematics we base our analysis on 
can be found in \cite{bemu2}.
The physical observables we use as examples are the single spin asymmetry (SSA) and the (unpolarized)  
azimuthal angle asymmetry (AAA) as well as the photon level cross section, 
which we will define below. We will demonstrate the feasibility of  
constraining  the GPDs from experimental data using these observables, as well as 
the relative size of the NLO corrections to the observables.  

For our input GPDs we use a modified form of the double distribution ansatz due 
to Radyushkin \cite{rad,rad2}, which assumes that the dependence on $t = (P-P')^2$ factorizes. 
In \cite{frmcbig} we demonstrated that the unmodified ansatz overshoots the DESY $e p$ collider HERA data, essentially because it 
samples an extrapolation of the inclusive unpolarized PDFs, which rise steeply as $x$ decreases,  
right down to $x=0$. For the quark distribution $q(x) = xq(x)/x \sim x^{-1.3}$, this leads 
to a large enhancement of the quark GPD relative to the PDF at the input scale.
Our modification is to introduce a lower cutoff on $x$, which we motivate below, and it 
leads to a good description of the available data. Like inclusive PDFs, 
the input GPDs are unknown nonperturbative boundary conditions which should ultimately be 
constrained by a QCD analysis of the data. The H1 data already appears to be ruling out 
Radyushkin's ansatz in its unmodified form.
NLO analyses of the evolution of GPDs and the DVCS amplitudes themselves have been carried 
out previously for large $x_{bj}$ \cite{pet,bemu1} and at LO for DVCS observables \cite{ffs,bemu3}. 
 
The GPD combinations, appearing in the representation of the factorization theorem \cite{rad,cf} 
for the DVCS amplitude used in our analysis, are the singlet-type combinations for each parton species, 
which are defined by 
\begin{align} 
&{\cal F}^{S(a),V/A} (X,\zeta,\mu^2,t) = \nonumber \\ 
&\left[\frac{H^{a,V/A} (v,\zeta,\mu^2,t) \mp H^{a,V/A} (-v,\zeta,\mu^2,t)}{(1-\zeta/2)}\right] \, , \nonumber \\ 
&{\cal F}^{g,V/A} (X,\zeta,\mu^2,t)    = \nonumber \\ 
&\left[\frac{H^{g,V/A} (v,\zeta,\mu^2,t) \pm H^{g,V/A} (-v,\zeta,\mu^2,t)}{(1-\zeta/2)}\right] \, ,  
\label{defsinglet} 
\end{align}
where $\zeta$ is the skewedness parameter, $v = (X -\zeta/2)/(1-\zeta/2)$, $\mu$ the factorization
scale, $a=u,d,s$ labels the quark flavor,  and $V$ stands for the
unpolarized (vector) case and $A$ for the polarized (axial-vector) case. 

Our representation of the GPDs is identical to the nondiagonal representation defined in \cite{golbier} and is different from the usual one (see for example \cite{pet}), which is defined symmetrically with respect to the incoming and outgoing nucleon plus momentum (defined on the interval $v \in [-1,1]$ and symmetric about $v=0$). The GPDs in Eq.\ (\ref{defsinglet}) have plus momentum fractions (on the interval $X \in [0,1]$) with respect to the incoming nucleon momentum, $P$, in analogy to the PDFs of inclusive reactions, with the ERBL region in the interval $X \in [0,\zeta]$ and the DGLAP region in the interval $X \in [\zeta,1]$. The transformation between the symmetric and nondiagonal representation is given in \cite{golbier}. Furthermore, within the nondiagonal representation $\zeta = x_{bj} = -q^2/ 2P \cdot q $. The symmetries of the GPDs, which were previously manifest about $v=0$, are now manifest about the point $X=\zeta/2$. Note that ${\cal F}^{S,V}$ and ${\cal F}^{g,A}$ are therefore antisymmetric and ${\cal F}^{S,A}$ and ${\cal F}^{g,V}$ are symmetric about the point $X=\zeta/2$.   
 
The DVCS amplitude, ${\cal T}_{DVCS} (\gamma^{*} p \to \gamma p)$, factorizes 
\cite{rad,cf} into a convolution of GPDs with coefficient functions: 
\begin{align} 
&{\cal T}^{S,V/A}_{DVCS} (\zeta,Q^2,t) = \sum_a e^2_a \frac{2 - \zeta}{\zeta}\Big[ \Big.\nonumber \\ 
&P.V. \int^1_0 dX~T^{S,V/A} \left(2X/\zeta - 1\right) {\cal F}^{S(a),V/A} (X,\zeta,Q^2,t) \nonumber\\ 
&\Big. \mp \int^1_{\zeta} dX~T^{S,V/A} \left(1 - 2X/\zeta\right){\cal F}^{S(a),V/A} (X,\zeta,Q^2,t)\Big] \, ,\nonumber\\ 
&{\cal T}^{g,V/A}_{DVCS} (\zeta,Q^2,t) = \frac{1}{N_f}\left (\frac{2 - \zeta}{\zeta}\right )^2\Big[ \Big.\nonumber \\ 
&P.V. \int^1_0 dX~T^{g,V/A} \left(2X/\zeta - 1\right){\cal F}^{g,V/A} (X,\zeta,Q^2,t)\nonumber\\ 
&\Big. \pm \int^1_{\zeta} dX~T^{g,V/A} \left(1 - 2X/\zeta\right){\cal F}^{g,V/A}(X,\zeta,Q^2,t)\Big] \, , 
\label{tdvcs} 
\end{align}
with the choice $\mu^2=Q^2=-q^2$ and the $T$'s are the 
LO or NLO coefficient functions taken from Eqs. (14) - (17) of \cite{bemu1}. 
$P.V.$ stands for the Cauchy principal value prescription, which implements 
the $+i\epsilon$ prescription to regulate the divergences in $T(2X/\zeta -1)$ at $X=\zeta$. 
Depending on the region of integration the coefficient functions have real and imaginary parts, which in 
turn generate real and imaginary parts of the DVCS amplitudes.
 
The triple differential cross section, on the lepton level, contains pure BH and DVCS terms, and an interference term. It is given in general by 
\begin{align} 
&\frac{d\sigma^{(3)} (e^{\pm} p \to e^{\pm} \gamma p)}{dx_{bj} dQ^2d|t|} = \int^{2\pi}_0d\phi \frac{d\sigma^{(4)}}{dx_{bj} dQ^2d|t|d\phi} \, = \nonumber \\ 
&\frac{\alpha_{e.m.}^3 x_{bj} y^2}{8\pi Q^4}\left(1+\frac{4M^2 x_{bj}^2}{Q^2}\right)^{-1/2} \int^{2\pi}_0d\phi|{\cal T^{\pm}}|^2 \, ,  
\label{crossx} 
\end{align}
\noindent where 
\bea 
&|{\cal T^{\pm}}|^2 = \nonumber 
&|{\cal T}_{DVCS}|^2 \pm ({\cal T}^*_{DVCS}{\cal T}_{BH} + {\cal T}_{DVCS} {\cal T}^*_{BH}) + |{\cal T}_{BH}|^2 \, ,   
\label{tdef} 
\eea 
and $\phi$ is the relative angle between the lepton and hadron scattering planes 
(we use the special Lorentz frame of \cite{bemu2} to define $\phi$), $y= Q^2 / x_{bj} S$ is the energy fraction 
lost by the incoming lepton, $S$ is the total center of mass energy and M is the nucleon mass. 
The exact, leading twist, expressions for the DVCS square, interference and BH square, 
both for an unpolarized and longitudinally polarized probe on an unpolarized target, were taken
from Eqs. (24) - (32) of \cite{bemu3} with the full $\phi$ dependence of the BH 
propagators restored \cite{frmcbig}. 

The (unpolarized) azimuthal angle asymmetry (AAA) and the single spin 
asymmetry (SSA) are defined by
 
\begin{align}
&\mbox{AAA} = \nonumber \\ 
&\frac{\int^{\pi/2}_{-\pi/2} d\phi (d\sigma^{(4)}-d\sigma^{BH}) - 
\int^{3\pi/2}_{\pi/2} d\phi (d\sigma^{(4)}-d\sigma^{BH})}{\int^{2\pi}_{0} d\phi d\sigma^{(4)}} \, , \label{aaadef} 
\end{align}
\noindent and 
\begin{align}
&\mbox{SSA} = 2 \frac{\int^{2\pi}_{0} d\phi \sin \phi \Delta\sigma^{(4)}}{\int^{2\pi}_{0} d\phi d\sigma^{(4)}} \, , 
\end{align}
where $\Delta\sigma^{(4)} = d\sigma^{\uparrow} - d\sigma^{\downarrow}$ with $\uparrow$ and $\downarrow$ signifying that the electron or positron beam is polarized along or against the beam direction, respectively. These definitions make the asymmetries directly proportional 
to the real part of a combination of DVCS amplitudes, for AAA, and the imaginary part of a 
combination of DVCS amplitudes, for SSA. 
For small $x$ and $t$, these combinations of amplitudes reduce to just the unpolarized helicity nonflip amplitude (which are obtained from a form factor decomposition of the $H$ functions defined above) for both the AAA and SSA \cite{bemu3,foot2}. 
 
As a model for our input GPDs we use the double distribution (DD) representation \cite{rad} with a 
factorized ansatz for the $t$-dependence \cite{rad2} of the DDs, which has been the model 
chosen for virtually all phenomenological studies of DVCS: 
\begin{align} 
&F_{DD}^{q(a)/g,V/A} (x',y',Q^2_{0},t) = \nonumber \\ 
&\pi^{q,g} (x',y') \, f^{q(a)/g,V/A} (x',Q^2_{0}) \, r^{q/g,V/A}(t) \label{DDs} 
\end{align} 
where the profile functions $\pi^{q,g} (x',y')$ are asymptotic shape functions \cite{rad} for quarks and gluons. 
The $f^{q(a)/g,V/A}$ are unpolarized/polarized inclusive input PDFs defined at the input scale $Q_0^2$.
The factorized $t$-dependence of the DDs is given by form factors, 
$r^{q,g,V/A} (t)$, depending on the parton species 
and whether the proton flips its helicity or not \cite{ji}. Here, we follow the model specified 
in Section.(4) of \cite{bemu3}, with a further assumption that the gluon distributions have the 
same $t$-dependence as the sea \cite{frmcbig}. 
The GPDs that we use for evolution are the nondiagonal ${\cal F}$s of Eq.(\ref{defsinglet}):  
thus to define the input model we require the reduction formula relating the $H(v)$-functions to the DDs: 
\begin{align} 
H (v,\zeta) = &\int^{1}_{-1} dx' \int^{1-|x'|}_{-1+|x'|} dy' \nonumber \\
              &\delta \left(x' + \frac{\zeta y'}{2-\zeta} - v\right) \, F_{DD} (x',y') \, . 
\label{reduction} 
\end{align} 
Performing the $y'$ integration, using the $\delta$ function, modifies the limits on the $x$ integral according to
the region. For example, in the DGLAP region, $X>\zeta$, one has for the quark distribution
\begin{align} 
&{\cal F}^{(a)} = \int^{X}_{\frac{X-\zeta}{1-\zeta}} dx' q^{(a)} (x') ~\pi^{q} (x',\frac{2X - \zeta - x' (2-\zeta)}{\zeta}) \, .   
\label{probint} 
\end{align} 
\noindent so that as $X \rightarrow \zeta$ the region $x' \rightarrow 0$ is probed in the PDF. 
For the polarized case this presents no problem since the available polarized distributions
tend to zero as $x' \rightarrow 0$ and this region is irrelevant to the integral. However, 
small $x$ inclusive unpolarized PDFs are known to increase rapidly as $x$ decreases 
so that this model is very sensitive to the extrapolation of the PDFs into the unmeasured region. 
Taking the model at face value leads to quark GPDs which are much larger than PDFs close to $X=\zeta$ 
and overshoots the HERA data on DVCS \cite{frmcbig}. This region is also probed in the ERBL region. 
 
However, if one considers the imaginary parts of particular DVCS graphs (which correspond to quarks and gluons 
on mass shell in the final state), one realizes that the limit $X \rightarrow \zeta$ cannot be realized in 
practice because the partons are obliged to produce finite mass hadrons in the final state. For example, for the process 
$g \gamma^* \rightarrow q {\bar q}$, we have $(X P + q)^2 = M^2_{q {\bar q}}$ which at the input scale, $Q_0$, 
implies $X - \zeta > M^2_{q {\bar q}} \zeta / Q_0^2 $. In fact one can argue that $X - \zeta > a \zeta$ for each of the 
graphs, where $a \sim {\cal O} (m^2_{hadron}/Q_0^2)$. In other words, finite mass effects become important for $X = \zeta$.
As a phenomenological model, we decide to pick the generic hadronic mass to equal the lightest vector meson mass,  
$m_{hadron} = m_{\rho} = 0.77$~GeV, which gives $a \approx 1/2$ for $Q_0 = 1$~GeV, and replace the lower limit 
of the integral in Eq.(\ref{probint}) (and all similar integrals) with $a \zeta \approx \zeta/2$. 
This change\cite{note2} ensures that the PDFs 
are only sampled at $x' \sim {\cal O} (\zeta) $ and above and thereby reduces the enhancement of the GPDs relative 
to the PDFs (see \cite{fmw} for further discussion).

In addition to these contributions from the double distributions the unpolarized singlet GPDs also contain a ``D term'' \cite{vanderhagen1,poly}, which is only nonzero in the ERBL region, and ensures the correct polynomiality in 
$\zeta$ \cite{poly} of the GPD and has a physical interpretation as a virtual meson exchange in the $t$ channel. 
The unpolarized helicity-flip GPD, due to lack of knowledge, is assumed to contain only the D term with an overall minus sign as required by the first moment sum rule for the sum of unpolarized helicity-flip and nonflip, in which the individual D terms must cancel. The polarized helicity-flip GPD was taken to be the asymptotic pion distribution amplitude due to the presence of the pion pole in this GPD, thus it does not evolve. 
  
Within this class of input model, we specify a particular input model for the GPDs by using the latest Martin-Roberts-Stirling-Thorne (MRST) LO and NLO 
sets \cite{mrst01} for the unpolarized PDFs (which have been tuned to the latest inclusive data from HERA) 
and the Gl\"{u}ck-Reya-Stratman-Vogelsang (GRSV) \cite{grsv} set for the polarized partons (for use in 
Eqs. (\ref{defsinglet}) (\ref{DDs}) (\ref{reduction}) ) at the input scale $Q^2_0 = 1~\mbox{GeV}^2$. 
We illustrated the spread of predictions one finds using different input PDFs in \cite{frmcbig}.
In the calculation of these observables, only gluons, up, down and strange quarks were used. 
The small charm quark contribution, generated by evolution, has justifiably been neglected 
so far (see \cite{frmc2} for more details). 

The input GPDs are evolved in both LO and NLO using an evolution program based on direct numerical integration, 
specifically written for this purpose \cite{frmc1}. The DVCS amplitudes are then computed via 
numerical integration, using LO or NLO coefficient functions in Eq.(\ref{tdvcs}) as appropriate 
and we chose the renormalization scale to be equal to the factorization scale, $Q^2 = \mu^2$. 
It turns out that the theoretical uncertainties due to scale variation are smaller than the 
uncertainties due to different inputs, therefore we will give our results without error bands 
for the scale dependence (see \cite{frmc2,frmcbig} for more details). 
The resultant DVCS amplitudes, calculated to both LO and NLO accuracy, using 
the respective LO and NLO input models, were the input for a further numerical program which computes the final 
DVCS observables (for details see \cite{frmcbig}). We will publish all of these codes on the 
Internet for use by the community \cite{website}. 

\begin{figure}  
\centering 
\mbox{\epsfig{file=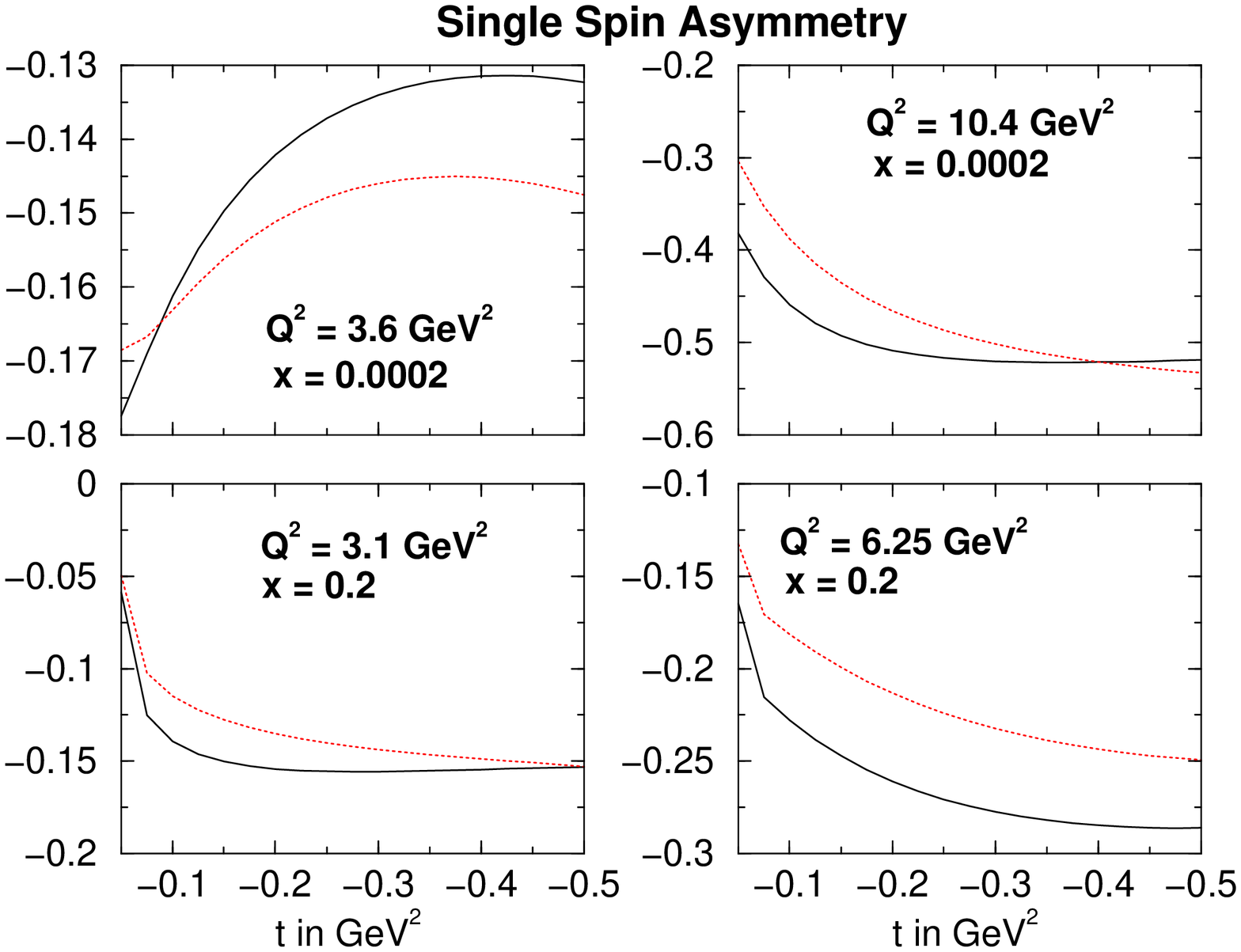,width=7.5cm,height=5.8cm}} 
\mbox{\epsfig{file=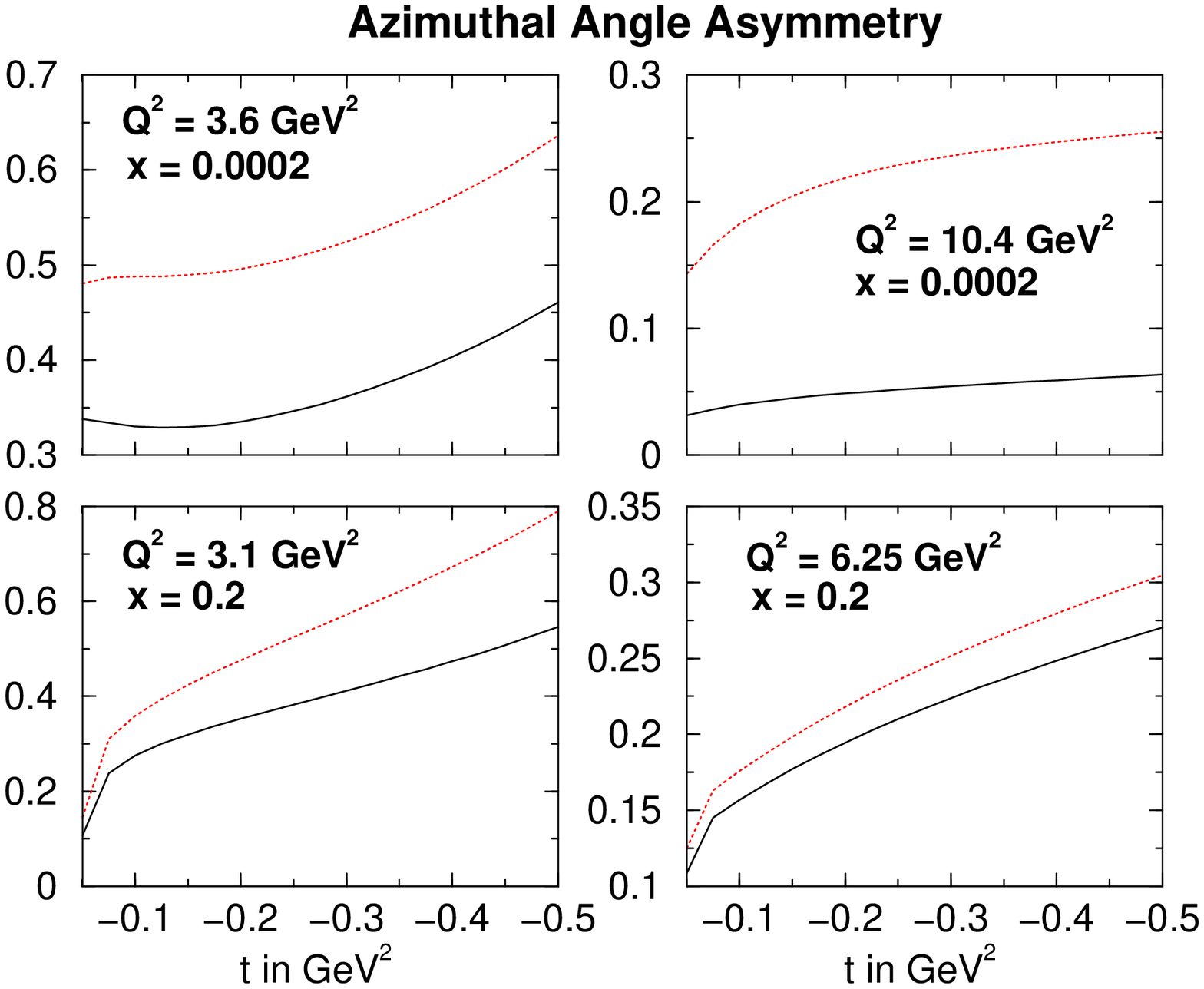,width=7.5cm,height=5.8cm}} 
\caption{The single spin and azimuthal angle asymmetries as functions of 
$t$ for ZEUS/H1 ($\zeta = x_{bj} =0.0002$) and HERMES ($\zeta = x_{bj} = 0.2$ ) kinematics, using MRST01 (unpolarized) and GRSV (polarized) input PDFs.
The solid curves are the full LO results and the dashed  
curves show the full NLO results for the respective input GPDs.}  
\label{fig:1} 
\end{figure}

In Fig. \ref{fig:1} we show representative NLO results for the 
azimuthal angle and single spin asymmetries, which by construction are 
directly sensitive to linear combinations of real and imaginary parts of DVCS amplitudes, 
for two values of the skewedness parameter, i.e. $\zeta = x_{bj} = 0.2, 0.0002$, 
which are indicative of values accessible at HERMES and ZEUS/H1, respectively. 
We note that the NLO predictions for both observables are 
discernibly different from the LO ones, although the NLO corrections to the 
single spin asymmetry are never more than about $20\%$, whereas the NLO 
corrections to the azimuthal angle asymmetry can be larger than $100\%$. 
This shows that the appearance of the gluon at NLO is stronger in the real part 
than in the imaginary part of the amplitude (see also \cite{frmc2}). 

The HERMES Collaboration \cite{exp} quotes two values for SSA: $SSA = -0.18 \pm 0.05 \pm 0.05$ and 
$\langle SSA \rangle = -0.23 \pm 0.03 \pm 0.04$. The first, which assumes that their missing mass equals the proton mass, 
is quoted at the following average values: $<\!x\!> = 0.11, <\!Q^2\!> = 2.5$~GeV$^2$ and $<\!t\!> = -0.27$~GeV$^2$. 
The second (average) value is the SSA integrated over the missing mass at the same average values of $x, Q^2$ and $t$. 
At this point in $x, Q^2, t$, we find $SSA= -0.31$ in LO and $SSA= -0.20$ in NLO for MRST01/GRSV00 input PDFs, i.e. 
good agreement. 

\begin{figure}  
\centering 
\mbox{\epsfig{file=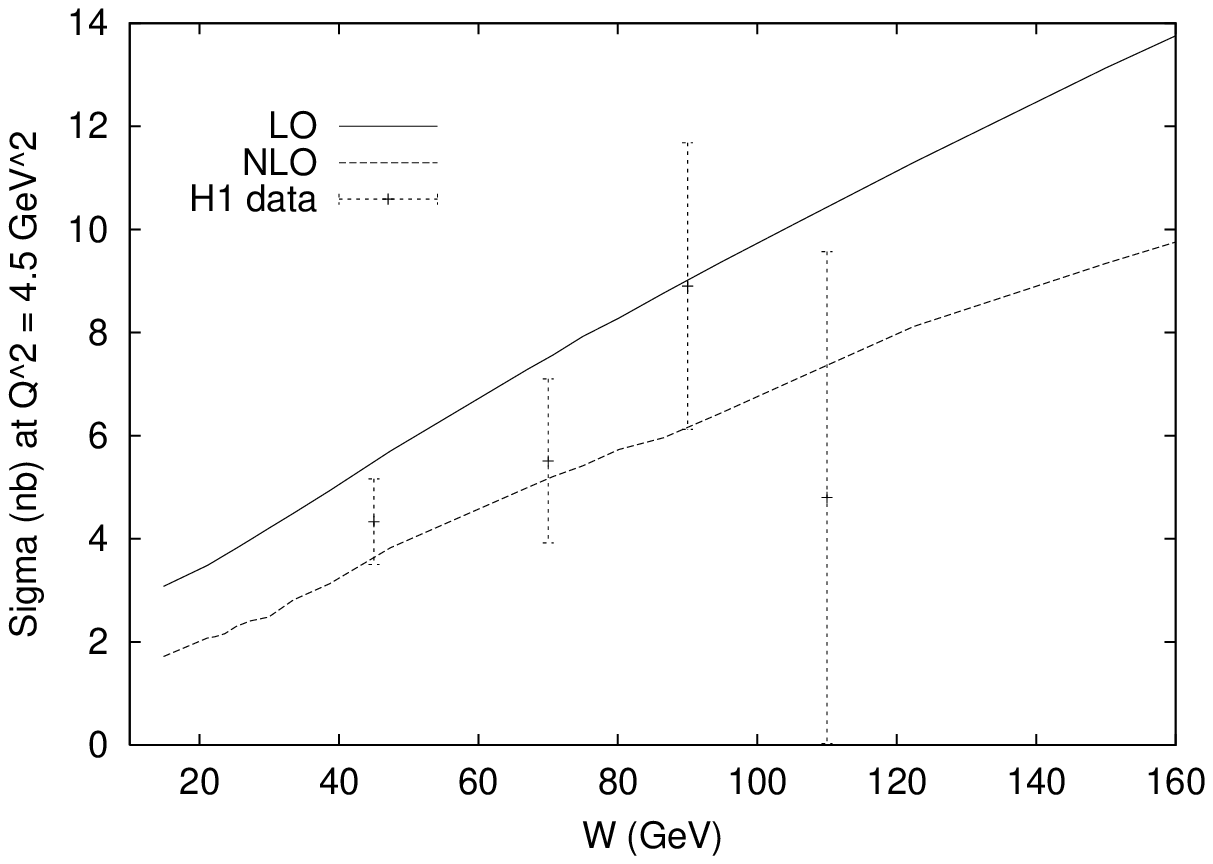,width=7.5cm,height=4.6cm}}
\mbox{\epsfig{file=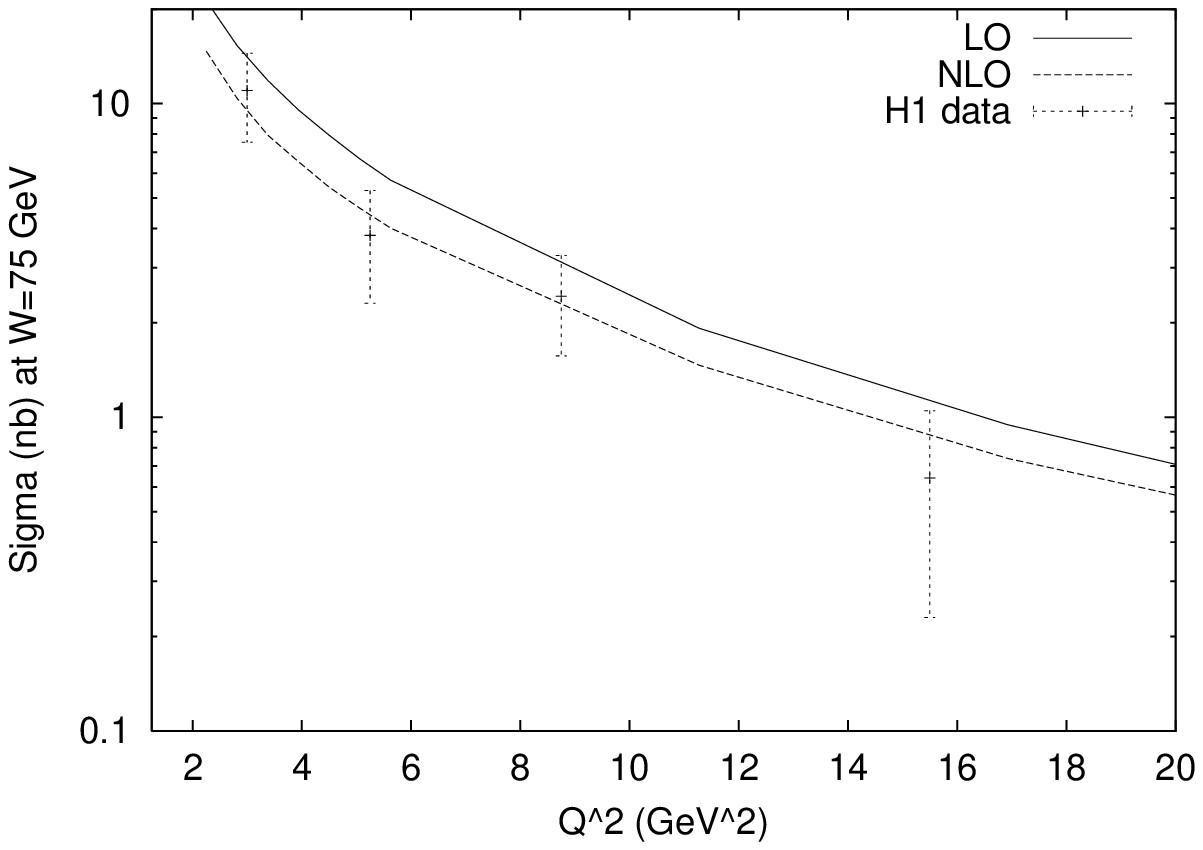,width=7.5cm,height=4.6cm}}
\caption{Photon level cross section $\sigma (\gamma^* P)$, at $Q^2 = 4.5$~GeV$^2$ in $W$ (upper figure) and at $W=75~\mbox{GeV}$ in $Q^2$ (lower figure), for MRST/GRSV at LO and NLO.}
\label{h1fig}
\end{figure}

We now compare our model to the recent H1 data on the $\sigma(\gamma^*p\to \gamma p)$ photon 
level cross section in Fig. \ref{h1fig}, defined through 
\begin{align} 
\sigma_{DVCS}(\gamma^*p \to \gamma p) = \frac{\alpha^2x^2\pi}{Q^4{\cal B}}|{\cal   T}_{DVCS}|^2|_{t=0}, 
\end{align} 
where ${\cal B}$ stems from the $t$ integration and, within our model for the 
$t$ dependence, was found to be about ${\cal B} \approx 6.5$~GeV$^{-2}$.
With our modification to Radyushkin's ansatz both LO and NLO curves appear to agree well with the data.  
Note the decrease of about $20-40\%$ in going from LO to NLO, which is due to a large, negative gluon amplitude 
and in line with results from inclusive reactions.

We have modified Radyushkin's ansatz for the GPDs by introducing a $\zeta$-dependent lower cutoff, 
$a \zeta \approx \zeta/2$, on the sampling of the PDFs at small $x$. 
Such a cutoff may be justified by arguing that the region $X \rightarrow \zeta$ is suppressed by 
finite mass effects. Without further tuning of $a \sim {\cal O} (m^2_{hadron}/Q_0^2)$, we can produce a good  
description of the current data using a NLO QCD model. As the precision of the data improves 
it may become necessary to use this cutoff as a fit parameter to constrain the input GPDs. 
We compare our NLO model to the LO result to quantify the size of the NLO QCD 
corrections to the DVCS observables.

We thank M. Diehl, D. M\"{u}ller and C. Weiss for helpful discussions. 
A.F. was supported by the DFG (FR 1524/1-1), and M.M. by PPARC.


\begin{thebibliography}{99}
\bibitem{foot1} Since the final state photon is elementary, one does not 
require additional non-perturbative information, as is the case in exclusive vector meson production.
\bibitem{foot2} This definition of the AAA is different from the usual one  
(see, e.g. the first reference of \cite{ffs}) and ensures that the numerator contains only the 
interference term and is thus directly proportional to a DVCS amplitude.
\bibitem{note2} Note that this modification doesn't change the normalization of the shape function 
or influence the forward limit, $\zeta\to 0$!
\end{thebibliography}
\end{document}